\documentclass[draft]{agujournal2019}
\usepackage{url} 
\usepackage{lineno}
\usepackage[inline]{trackchanges} 
\usepackage{soul}
\draftfalse

\journalname{JGR: Planets}

\begin{document}

%
%


\title{Onset of dynamo action in planetesimals}

%
%




\authors{Ludovic Huguet\affil{1}, Jonathan E. Mound\affil{1}, Christopher J. Davies\affil{1}, James F.J. Bryson\affil{2} }


\affiliation{1}{School of Earth and Environment, University of Leeds, Leeds, LS2 9JT, UK}
\affiliation{2}{Department of Earth Sciences, University of Oxford, Oxford OX1 3AN, UK}




\correspondingauthor{Ludovic Huguet}{l.g.huguet@leeds.ac.uk}



\begin{keypoints}
\item We study the onset of dynamo action in planetesimal-like conditions, using direct numerical simulations.
\item Extrapolation to planetesimal conditions sets the threshold for dynamo action as at least 1000 times the onset for convection.
\item This criterion extends the likely duration of magnetic field generation and lowers the minimum core radius required for dynamo action.
\end{keypoints}

%
%

%
%

\begin{abstract}
Several meteorites have been found to carry primary remanent magnetizations imparted by fields generated within planetesimal cores during the early solar system. Thermal evolution models have shown that thermal convection likely drove the dynamo in the early evolution of these bodies. In such small cores, the magnetic Reynolds number is thought to be close to the threshold for dynamo action. However, the critical value of the magnetic Reynolds number, meaning its value at the onset, is also poorly constrained in dynamo simulations. We perform dynamo simulations to investigate the onset of the dynamo at different Ekman ($E$) and magnetic Prandtl ($Pm$) numbers, in a quasi-full sphere. Along two empirical paths in this $(E,Pm)$ space, the onset of dynamo action depends on whether the magnetic field is initially strong or weak. The onset of the dynamo occurs at larger supercriticality (ratio between the Rayleigh number ($Ra$) and its critical value ($Ra_c$)) when moving toward more realistic parameter values. Once extrapolated to planetesimal core conditions, the supercriticality for the onset of the dynamo is about $10^3$ for a strong initial magnetic field (meaning magnetic energy is of the same order of magnitude or larger than the kinetic energy). For a weak initial field, the required $Ra/Ra_c$ would be larger. Compared to previous estimates, this revised criterion facilitates dynamo activity, and so extends the lifetime of the magnetic field compared to previous models and allows the generation of a magnetic field in smaller planetesimals.
\end{abstract}
\section*{Plain Language Summary}
Many meteorites preserve magnetic signatures acquired in the early solar system. These signatures can have been produced by magnetic fields generated inside molten metallic cores of small bodies; however, it remains uncertain whether, or for how long, a given planetesimal can sustain dynamo action. We use computer simulations to determine the conditions required for magnetic field generation and extrapolate our results to planetesimal core conditions. We find that the threshold for dynamo action can be expressed in terms of the vigour of thermal convection. Under planetesimal conditions, sustaining a magnetic field requires convection to be roughly a thousand times stronger than the minimum required for fluid motion. The required threshold also depends on whether a strong external magnetic field already exists. Our revised criterion suggests that magnetic fields could have been generated in smaller planetesimals than previously thought, helping explain the magnetizations recorded by ancient meteorites.

\section{Introduction}
Since the first palaeomagnetic measurements on meteorites in the 1950s (\citeA <see>{weiss2010paleomagnetic} for a review), advances in the field have made it possible to distinguish whether the recorded magnetization reflects a terrestrial or an extraterrestrial magnetic field \cite{weiss2008magnetism}, reliably determine the nature of the field that imparted the remanence, accurately recover palaeointensities, and dissect complex and multifaceted remanence acquisition processes. As a result, palaeomagnetism has emerged as a key method for investigating the magnetic and thermal evolution of early Solar System bodies.  
Today, dozens of meteoritic samples have been examined, revealing a broad spectrum of magnetic field strengths and time-dependent behaviours \cite <e.g.,>{carporzen2011magnetic,tarduno2012evidence,bwhhk2017}. These findings indicate that magnetic fields were widespread during the evolution of planetesimals.

Magnetic fields generated by the protoplanetary disk result from the motion of ionised gas within the solar nebula. Palaeointensity estimates from meteorites that formed a few million years after calcium–aluminium-rich inclusions (CAIs) indicate field strengths of $\mathcal{O}(10)~\mu\mathrm{T}$ during the first $\sim$2 Myr, followed by a decrease to values of $\mathcal{O}(0.1)~\mu\mathrm{T}$ at later times ($~4-5$ Myr after CAI formation) \cite{weiss2021history}. For reference, the present-day geomagnetic field intensity at Earth’s surface is approximately $30-60~\mu\mathrm{T}$. While the earliest signals of a magnetic field are consequently attributed to nebular magnetism, the persistence of magnetic remanence at later times can be explained by dynamo action within the cores of differentiated planetesimals \cite{bnn2019,dodds2021thermal,sbn2024}. 

In modelling studies of the thermal and chemical evolution of planetary cores, the existence of a dynamo-generated magnetic field is commonly inferred from the value of the magnetic Reynolds number, $Rm$ \cite{s2010b,scheinberg2015magnetic,rbs2015,kv2018,bnn2019,dodds2021thermal,zr2022}. The magnetic Reynolds number quantifies the competition between inductive processes (stretching and advection of the magnetic field by a flow of typical velocity) and magnetic diffusion. Parameterisations of coupled core-mantle dynamics yield the time evolution of the convective power in the core (derived from a thermal or compositional buoyancy flux), which can be related to $Rm$ by an assumed scaling law. The viability of dynamo action is assessed by comparing this inferred $Rm$ (which is a function of time) to the critical value, which is estimated from theory or numerical studies of dynamo action. For small planetary bodies, previous thermal evolution and dynamo generation models find $Rm< 100$ \cite<e.g.,>{sbn2024}, which is comparable to existing estimates for the critical value \cite<e.g. $Rm \sim 40$ in>{ca2006}. 

The calculation of $Rm$ requires an estimate of the typical core flow speed. Using numerical simulations, \citeA{alp2009} demonstrated that the typical flow velocity, as characterised by the Rossby number (ratio between inertia and Coriolis forces), scales as a power law of the convective power, consistent with theoretical predictions \cite{c2010,d2013}. However, the prefactor of such scaling laws is often not very well constrained, as these laws likely encompass extra dependencies not taken into account in the main power law (e.g., diffusivities, boundary conditions, driving forces, size of the inner core; \citeA{ca2006,c2010,d2013}). From the scaling law between the Rossby number and the convective power, a typical velocity scale can be determined, and so can the magnetic Reynolds number. However, the uncertainty around the prefactor leads to an uncertainty of about $50\%$ in $Rm$ (\citeA{alp2009} and this study). This uncertainty is significant because thermal evolution models predict a value of $Rm$ close to the threshold for dynamo action \cite{sbn2024,sbnd2025}.

Estimating the critical magnetic Reynolds number is challenging because the induction equation that governs magnetic field generation does not admit a unique critical value (the linearised operator is non-normal and no single mode governs the instability \cite{p2015}). This situation contrasts with the onset of non-magnetic convection, where a single critical value of the Rayleigh number determines the onset of instability \cite{chandrasekhar2013hydrodynamic}. Theoretical bounds on $Rm$ are derived from inequalities and energy estimates applied directly to the induction equation. This leads to estimates of the critical magnetic Reynolds number in the range of $1-10$ depending on the specific definition of $Rm$ \cite{b1958, b1975, c1969, p1977, p1979, chlllj2018}. These bounds are universal in the sense that they apply to all admissible velocity fields. Still, they are necessarily conservative and can significantly underestimate the dynamo threshold  \cite{p1979, p2015, chlllj2018}. By contrast, optimal bounds are obtained by identifying, analytically or numerically, classes of flows that minimise the magnetic Reynolds number required for sustained magnetic growth. These approaches yield substantially higher estimates of the dynamo threshold \cite{w2012,chj2015,chlllj2018,lcll2020}. Estimates of the minimal magnetic Reynolds number can also be obtained from direct numerical simulations of dynamo action. A commonly cited order-of-magnitude estimate of $Rm$ is $\mathcal{O}(60)$ \cite{ca2006,lfncs2022}.

The uncertainty of the scaling law for the magnetic Reynolds number as a function of the convective power, combined with the non-trivial choice of its critical value, places fundamental limitations on current estimates of the viability and longevity of dynamo action in small planetary bodies. In this paper, we seek to improve constraints on both of these quantities using numerical dynamo simulations. We also investigate the supercriticality of convection as an alternative proxy for quantifying the onset of dynamo action. This quantity has the advantage that it is a direct output from parameterised planetesimal evolution models and does not need to be inferred from a scaling law.

Direct numerical simulations of dynamo action in rotating spherical geometry cannot currently be undertaken at the physical conditions that characterise geophysical and astrophysical bodies, and so a strategy is needed to extrapolate from results obtained in numerically tractable parameter regimes. For Earth’s core, the key properties of the magnetic field and core flow are fairly well established: the field is predominantly dipolar with intermittent polarity reversals, and the core dynamics are governed by a Magneto-Archimedes-Coriolis balance, in which magnetic energy exceeds kinetic energy and the magnetic Reynolds number is $\mathcal{O}(10^3)$ \cite{d2013}. Building on this, \citeA{agf2017} proposed a pathway linking numerically feasible simulations to the parameter regime relevant to Earth’s core. They constructed a physically motivated trajectory in which the Ekman (ratio between viscous force and Coriolis force) and magnetic Prandtl (ratio between viscosity and magnetic diffusion) numbers are reduced in a controlled manner so that essential characteristics of Earth’s core are preserved in the simulations \cite{d2016,agf2017,a2019,dormy2025rapidly}. In contrast, analogous constraints on the magnetic field morphology and core flow are not available for planetesimals, for which we can generally only infer the existence and magnitude of a magnetic field. To study the onset conditions for dynamo action and the resulting magnetic field under planetesimal-like conditions, we adopt an empirical approach. Specifically, we define two distinct parameter-space paths that converge to the same endpoint representative of planetesimal conditions ($E = 10^{-13}$ and $Pm = 10^{-6}$), but originate from different magnetic Prandtl numbers (and from the same Ekman number). We explore approximately $30\%$ of each path, determining the critical conditions for dynamo onset and characterising the corresponding magnetic field and core flow at that threshold. We will demonstrate that certain diagnostic parameters can be extrapolated asymptotically to predict whether a planetesimal can sustain a magnetic field.

The paper is organised as follows. Section~\ref{sec:method} introduces the dimensionless governing equations, the dimensionless governing parameters, and the boundary conditions used to simulate a thermally driven dynamo in a spherical shell. We describe the numerical simulations, the range of dimensionless parameters used, and the main diagnostic parameters of the dynamo simulations. Section \ref{sec:results} presents our simulations in a dynamo regime diagram and the nature of the dynamo produced at onset along both paths. Using our dynamo simulations, we present a scaling law for typical velocity with convective power. Finally, we extrapolate the onset of dynamo action to planetesimal conditions. In Section~\ref{sec:discussion}, we discuss the implications for the generation of a magnetic field by thermally-driven convection in a planetesimal core.

\section{Method}
\label{sec:method}
\subsection{Governing equations}
 We consider the dimensionless Boussinesq magnetohydrodynamic equations in a rotating spherical shell in spherical coordinates ($r$, $\theta$, $\phi$). The system is nondimensionalised using the outer core radius $r_o$ as the length scale, the magnetic diffusion time $\tau_\eta=r_o^2/\eta$ as the time scale, and the magnetic field scale $B_0 = (\rho \mu_0 \eta \Omega)^{1/2}$, with $\eta$ the magnetic diffusivity, $\mu_0$ the magnetic permeability, $\rho$ the density, and $\Omega$ the rotation rate. Gravity varies linearly with radius as $\mathbf{g} = -g_o (r/r_o)\mathbf{\hat{r}}$ where $g_o$ is the acceleration of gravity at $r_o$.  The unit vectors aligned with the rotation axis and the spherically radial direction are $\mathbf{\hat{z}}$ and $\mathbf{\hat{r}}$, respectively. We solve the induction equation for the magnetic field $\mathbf{B}$, the momentum equation for the velocity $\mathbf{u}$, and the temperature equation for the temperature $T$. The temperature equation is nondimensionalised using $\beta r_o^2$, where $\beta = \hat{S}/(3q)$ is defined from the conductive base state of the heat equation using an internal heating source (see Section~1.1 in Supporting Information). With this choice, the dimensionless volumetric heat source $S$ is equal to 3 in the governing equations. The Roberts number, $q=\kappa/\eta$, denotes the ratio of thermal to magnetic diffusivity. Under the Boussinesq approximation and in a rotating frame, the dimensionless induction, momentum, and heat equations take the form
\begin{eqnarray}
 \frac{\partial \mathbf{B}}{\partial t} -\nabla \times \left(\mathbf{u}\times\mathbf{B}\right) = \nabla^2\mathbf{B}, \label{eq:induction}\\
 \frac{\partial \mathbf{u}}{\partial t} + \mathbf{u}\times \left(\nabla \times \mathbf{u}\right) + \frac{2Pm}{E}\hat{\mathbf{z}}\times \mathbf{u} &= -\nabla \mathcal{P}+Pm\nabla^2\mathbf{u} 
 - \frac{2 Pm^2 Ra}{Pr E} T \mathbf{r}
+ \frac{2Pm}{E}\left(\nabla\times \mathbf{B}\right)\times\mathbf{B}, \label{eq:momentum}\\
 \frac{Pr}{Pm}\left(\frac{\partial T}{\partial t} + \mathbf{u}\cdot \nabla T\right) =\nabla^2 T + S,\label{eq:heat}     
\end{eqnarray}
where $\mathcal{P}$ is the pressure. The incompressibility of the fluid and the divergence-free nature of the magnetic field imply that $\nabla \cdot \mathbf{u} = 0$ and 
$\nabla \cdot \mathbf{B} = 0$. 
The dynamics of the system are controlled by four dimensionless parameters:
\begin{eqnarray*}
\mathrm{Ekman~number:}& \quad
E = \frac{\nu}{\Omega r_o^2}, \quad
&\mathrm{Rayleigh~number:} \quad
Ra = \frac{\alpha g_0 \beta r_o^4}{\nu \kappa},\\
\mathrm{Prandtl~number:}& \quad
Pr = \frac{\nu}{\kappa}, \quad
&\mathrm{Magnetic~Prandtl~number:} \quad
Pm = \frac{\nu}{\eta}.
\end{eqnarray*}
where $\nu$ is the kinematic viscosity, $\kappa$ is thermal diffusivity, and $\alpha$ is the thermal expansion coefficient.

In the presence of an inner core, a natural alternative for the characteristic length scale is the spherical-shell thickness, $D = r_o - r_i = r_o(1-\chi)$, where $\chi = r_i / r_o$ is the ratio of inner to outer core radii. This choice of scaling is commonly adopted in Earth-like simulations and is also relevant here, as our simulations include a small inner core ($\chi = 0.01$) employed to avoid the singularity at the origin of the spherical coordinate system. Numerous investigations using either a small inner core or a full sphere exhibit qualitatively similar dynamics \cite{alp2009,la2011,clns2019,blmj2025}. We also reduce the influence of our inner core through the choice of boundary conditions. The magnetic field satisfies insulating boundary conditions at both the inner and outer boundaries. The velocity field is subject to no-slip conditions at the outer boundary and free-slip conditions at the inner boundary. For the temperature field, fixed-flux boundary conditions are imposed at both boundaries, with zero heat flux at the inner boundary. Internal heat sources are distributed uniformly throughout the fluid shell, consistent with full-sphere internally heated convection. The Prandtl number is fixed to $Pr=1$ throughout this study.

\subsection{Empirical Paths}
\label{subsec:path}

To explore dynamo onset across a wide range of parameters, we define two empirical paths in $(E,Pm)$ parameter space. Both paths are constructed to extrapolate from numerically accessible values to those expected for a planetesimal core, with target parameters $E^A \sim 10^{-13}$ and $Pm^A \sim 10^{-6}$ \cite{sbnd2025}. The two paths originate from a common Ekman number $E_0 = 2 \times 10^{-4}$ but from different initial magnetic Prandtl numbers, $Pm_0^{(1)} = 80$ and $Pm_0^{(2)} = 3$.

We introduce a path parameter $\epsilon = E/E_0$, which measures progression along each path. By construction, $\epsilon = 1$ corresponds to the initial numerical models, while $\epsilon \simeq 5 \times 10^{-10}$ corresponds to planetesimal-like conditions. For each path, the relationship between $\epsilon$ and $Pm$ is prescribed as a power law,
\begin{equation}
Pm^{(i)} = Pm_0^{(i)} \epsilon^{n^{(i)}},
\end{equation}
where the exponent $n^{(i)}$ is chosen such that both paths terminate at $(E^A,Pm^A)$. This requirement yields
\begin{equation}
n^{(i)} = \frac{\log(Pm^A/Pm_0^{(i)})}{\log(E^A/E_0)}.
\end{equation}
The resulting empirical paths are therefore given by
\begin{eqnarray}
E^{(1)} = E^{(2)} = \epsilon E_0, \qquad
Pm^{(1)} = \epsilon^{0.8497} Pm_0^{(1)}, \qquad
Pm^{(2)} = \epsilon^{0.6965} Pm_0^{(2)}.
\end{eqnarray}

Path~2 closely follows the expected scaling of the critical magnetic Prandtl number inferred from numerical fits and theoretical arguments, namely $Pm_c \propto E^{3/4}$ or $Pm_c~\propto~E^{2/3}$ \cite{ca2006,dl2008}. Path~1, which starts from a larger initial value of $Pm$, reaches the same endpoint and therefore follows a slightly steeper slope in $(E,Pm)$ space.

\begin{figure}
 \includegraphics[width=\textwidth]{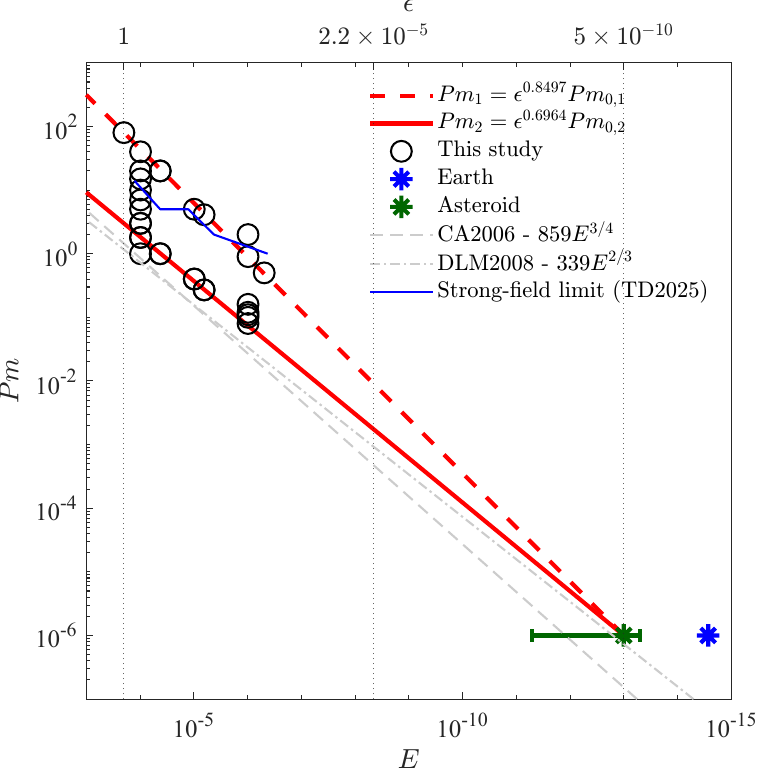}
 \caption{Exploration of magnetic Prandtl number $Pm$ and Ekman number $E$ parameter space. Empirical paths 1 (red dashed line) and 2 (red solid line) are indicated. Black circles show simulations performed in this study that sustain a magnetic field by dynamo action. The blue and green stars are the typical values of the Ekman and magnetic Prandtl numbers for the Earth's core and a planetesimal (or planetesimal) core, respectively. Grey dashed, and dash-dotted lines are scaling laws for $Pm$ based on best fit or theoretical arguments \cite{ca2006,dl2008} (CA2006, DLM2008 respectively). The blue line corresponds to the lower bound for strong-field dynamo as described by \cite{teed2025scaling} (TD2025). Vertical dotted lines denote the Ekman values of the path at the starting point $\epsilon = 1$, at $50\%$ of the path ($\epsilon = 2.2 \times 10^{-5}$), and at the planetesimal's parameters ($\epsilon = 5 \times 10^{-10}$).}
 \label{fig:PmE}
\end{figure}

Figure~\ref{fig:PmE} illustrates the explored parameter space, showing the two empirical paths together with the locations of the simulations performed in this study. For reference, typical values of $(E,Pm)$ for Earth’s core and for a planetesimal core are also indicated. We also indicate the suggested scaling laws for the minimal magnetic Prandtl number required for accessing a dipolar-dominated dynamo regime $Pm_c$ \cite{ca2006,dl2008}. We note that simulations along path 1 are initially in the strong-field dipolar regime, where the magnetic field energy is much larger than the kinetic energy ($\frac{E_B}{E_K}\gg10$) and $\Lambda>10$ \cite{dormy2025rapidly,teed2025scaling}, before moving to the weak-field dipolar regime. Path 2 has a smaller magnetic Prandtl number and is therefore always in the weak-field region.

\subsection{Supercriticality} \label{Rsection}

We define the supercriticality as the ratio of the Rayleigh number to its critical value for the onset of non-magnetic convection, at a given Ekman number
\begin{equation}
\mathcal{R} = \frac{Ra}{Ra_c}.
\label{eq:supercriticality}
\end{equation}
For each $(E,Pm)$ pair, the onset of dynamo action is investigated by running a suite of simulations with different values of $\mathcal{R}$. The critical Rayleigh number $Ra_c$ and the associated most unstable mode are computed using the open-source eigenvalue solver \textit{Kore}. This code has previously been used to determine $Ra_c$ in closely related configurations \cite{btcsa2023} and has been benchmarked against asymptotic linear stability analyses \cite{dsjjc2004}. We determine $Ra_c$ for the same geometric and thermal configuration as our direct numerical simulations (i.e., internal heating with fixed fluxes and a small inner core). 

\begin{figure}
 \includegraphics[width=\textwidth]{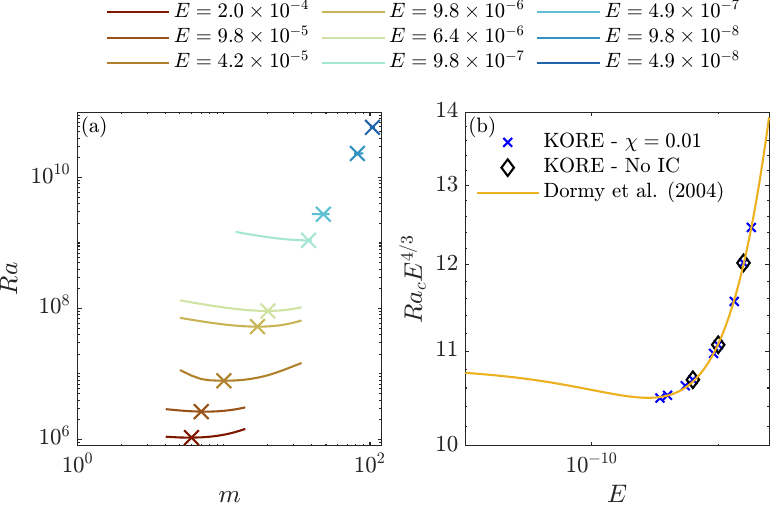}
 \caption{(a) Rayleigh number $Ra$ as function of the azimuthal mode $m$ for different $E$. Crosses indicate the most unstable mode, corresponding to the critical Rayleigh number $Ra_c$ for a small inner core. (b) $Ra_c E^{4/3}$ as a function of $E$. The blue crosses and black diamonds are the critical Rayleigh numbers for each given Ekman number with a small inner core and no inner core, respectively. The yellow line denotes the best fit of the leading order scaling laws from the asymptotic theory for the onset of convection \cite{dsjjc2004}. }
 \label{fig:rac_m_ra_E}
\end{figure}

Figure~\ref{fig:rac_m_ra_E}a shows the Rayleigh number as a function of the azimuthal wavenumber $m$ for nine Ekman numbers. For each curve, the minimum defines the critical Rayleigh number $Ra_c$ and the most unstable mode. These values are very close to those obtained in full-sphere configurations (Figure~\ref{fig:rac_m_ra_E}b). This similarity arises because internally heated convection tends to initiate instability within the bulk of the fluid rather than near the inner boundary, in contrast to differentially heated configurations, where instability is localised near the bottom boundary \cite{jsm2000,btcsa2023,blmj2025}. The resulting critical values are reported in Table~SI~1 in the Supporting Information. Figure~\ref{fig:rac_m_ra_E}b presents the variation of the scaled critical Rayleigh number $Ra_c E^{4/3}$ as a function of $E$. Results obtained with a small inner core and with no inner core are shown for comparison. The yellow line corresponds to the leading-order asymptotic scaling for the onset of convection under internal heating \cite{dsjjc2004}. In the asymptotic limit $E \ll 1$, both configurations follow the same scaling
\begin{equation}
 Ra_c \simeq 10.8 E^{-4/3}.
\end{equation}

\subsection{Diagnostic parameters}

We define a set of diagnostic quantities derived from the simulation outputs to characterise the kinetic and magnetic regimes explored. These global measures allow systematic comparison of simulations as a function of the input parameters.

The dimensionless kinetic and magnetic energies (scaled by $\rho \eta^{2} r_o$) are defined as
\begin{equation}
E_K = \frac{1}{2} \int \mathbf{u}^2 \, dV
\end{equation}
and
\begin{equation}
E_B = \frac{Pm}{E} \int \mathbf{B}^2 \, dV.
\end{equation}
The ratio of magnetic to kinetic energy is denoted by $M = E_B/E_K$. The magnetic Reynolds number and the Rossby number are computed from the kinetic energy:
\begin{equation}
Rm = \sqrt{\frac{2E_K}{V}} \quad \mathrm{and} \quad Ro = \frac{E}{Pm} \sqrt{\frac{2E_K}{V}}.
\end{equation}
To quantify the strength of the magnetic field, we use the Elsasser number, which is defined by the ratio between the Lorentz force and the Coriolis force as
\begin{equation}
 \Lambda = \frac{\tilde{\mathbf{B}}^2}{2\Omega\rho\mu_0\eta} = \frac{E}{Pm}\frac{E_B}{V},
 \label{eq:ell}
\end{equation}
where $\mathbf{\tilde{B}}$ is the dimensional magnetic field and $V\propto r_o^3$ is the fluid volume. 

To characterise the morphology of the magnetic field (dipolar or multipolar), we use the dipolar fraction $f_{dip}$, which is the ratio between the dipole component and the total magnetic field up to spherical harmonic degree $l_{\mathrm{max}}=12$, and is given by 
\begin{equation}
 f_\mathrm{dip} = \left( \frac
{\int{ \mathbf{B}_{l=1}(r=r_o) \cdot \mathbf{B}_{l=1}(r=r_o) }dS }
{\int{ \mathbf{B}_{l\leq 12}(r=r_o) \cdot \mathbf{B}_{l\leq 12}(r=r_o) }dS} 
\right)^{1/2},
\label{eq:fdip}
\end{equation}
with $S$ the spherical surface at $r_o$. Dynamos with $f_{\mathrm{dip}} > 0.5$ are classified as dipole-dominated, while those with $f_{\mathrm{dip}} \leq 0.5$ are multipolar.

Assuming a well-mixed core, the dimensionless convective power is defined as \cite{alp2009}
\begin{equation}
p = \frac{\Phi}{V} ,
\end{equation}
with $\Phi$ the dissipation nondimensionalised with $\rho \Omega^3 r_0^2$. For a statistically steady state, the time-averaged changes in kinetic and magnetic energies are negligible. Then, the total dissipation $\Phi$ is given by the sum of viscous and ohmic dissipation:
\begin{equation}
\Phi = \left( \frac{E}{Pm} \right)^3 \left[ Pm\int\left(\nabla \times \mathbf{u}\right)^2 dV + \frac{2 Pm}{E}\int\left(\nabla \times \mathbf{B}\right)^2 dV \right] .
\end{equation} 

\subsection{Scaling laws}

For saturated dynamos, the dominant force balance determines the velocity scaling. A magnetostrophic (MAC) balance between Lorentz, buoyancy, and Coriolis forces predicts $u \propto p^{4/9} \Omega r_o$ \cite{d2013}, while a balance between Coriolis, inertia, and buoyancy (CIA) yields $u \propto p^{2/5} \Omega r_o$ \cite{c2010}. Using the relationship between convective power and modified flux-based Rayleigh number ($p=\gamma Ra_Q$) \cite{alp2009}, the magnetic Reynolds number can be expressed in terms of supercriticality. For the CIA balance,
\begin{eqnarray}
Rm_{\mathrm{CIA}} &=
c_1 \gamma^{2/5} 10.8^{2/5}
\, Pm\, Pr^{-4/5} E^{-1/3}
\left( \frac{Ra_F}{Ra_c} \right)^{2/5},
\label{eq:rm_CIA}
\end{eqnarray}
whereas the MAC balance gives
\begin{eqnarray}
Rm_{\mathrm{MAC}} &=
c_2 \gamma^{4/9} 10.8^{4/9}
\, Pm\, Pr^{-8/9} E^{-7/27}
\left( \frac{Ra_F}{Ra_c} \right)^{4/9}.
\label{eq:rm_MAC}
\end{eqnarray}
Here $\gamma\sim0.5881$ is a constant describing the distribution of heat in a well-mixed outer core (for internal heating and a small inner core) \cite{alp2009}. $c_1$ and $c_2$ are proportionality pre-factors in $Ro = c_1 p^{2/5}$ or $Ro= c_2 p^{4/9}$, which can be inferred from the best fit of our simulations (see Figure~\ref{fig:Ro_p_Ro_p}, and with $Ra_c \simeq 10.8E^{-4/3}$ for internally heated convection). $Ra_F=Ra_Q \frac{Pr^2}{E^3}$ is the flux-based Rayleigh number as defined by \citeA{alp2009}. 

\subsection{Simulations}
Equations (\ref{eq:induction})–(\ref{eq:heat}) are solved using the Leeds Dynamo Code, which has been extensively benchmarked against other numerical dynamo models in a variety of configurations \cite{wsg2007,Mound2017HeatConditions,Matsui2016PerformanceModel}. The numerical resolution is chosen such that both kinetic and magnetic energy spectra exhibit at least a two-order-of-magnitude decay between their peak values and the spectral truncation at maximum spherical harmonic degree $l_{\max}$ and azimuthal order $m_{\max}$.

We explore the effects of varying the Ekman number $E$, the magnetic Prandtl number $Pm$, and the degree of supercriticality $\mathcal{R}$. For nearly every triplet $(E,Pm,\mathcal{R})$, two simulations are performed: one initialised with a saturated magnetic field, called strong-initial-field (SiF), corresponding to a magnetic-to-kinetic energy ratio $M>1$ (74 simulations), and one initialised with a weak magnetic field, called weak-initial-field (WiF), corresponding to $M<0.1$ (68 simulations). 

We ran 142 simulations and classified them into two groups: (i) dynamo-active if the magnetic energy remains statistically steady over sufficiently long integration times (66 simulations) or (ii) non-dynamo if the magnetic energy decays almost linearly in log-space or by several orders of magnitude (76 simulations). Our simulations span the parameter ranges $E = 2 \times 10^{-4}$ to $5 \times 10^{-7}$, $Pm = 0.05$ to $80$, and $\mathcal{R} \simeq 1.2$ to $310$. Most of the simulations (101) were set up to follow two empirical paths in $(E, Pm)$ parameter space (Section~\ref{subsec:path}); the remaining 41 explore off-path configurations. All simulations are integrated for several advection times beyond the initial transient, ensuring that a statistically saturated state is reached.

\section{Results}
\label{sec:results}
In this section, we will present the results from the dynamo simulations run along paths 1 and 2, as well as some off-path results.

\subsection{Dynamo regimes}
\begin{figure}
 \includegraphics[width=\textwidth]{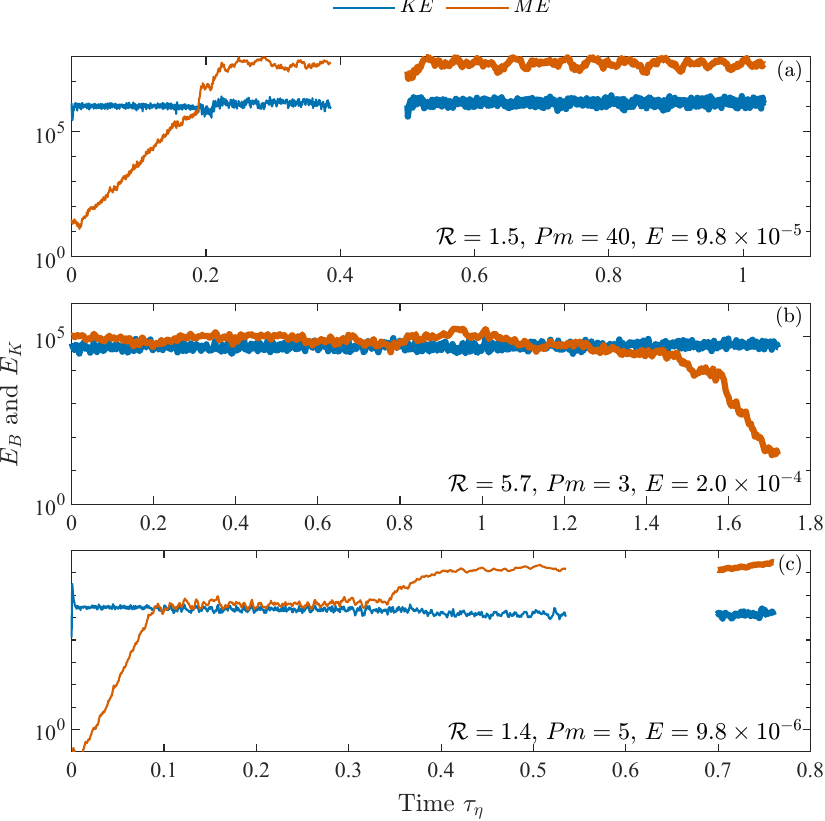}
 \caption{Evolution of the kinetic (blue) and magnetic (red) energies for five simulations with three different sets of input parameters. For each case (except (b)), simulations start with a weak-initial field (thin lines) or a strong-initial-field (thick lines), respectively. (a,c) A growing dynamo from a WiF and SiF. Both solutions end in the same saturated state. (b) Dying dynamo initiated with a strong-initial-field (SiF) after $1.5\tau_\eta$ with parameters along the path 2.}
 \label{fig:time_evolution}
\end{figure}

We have performed simulations to test whether a weak initial magnetic field ($M<0.1$ and $\Lambda<10^{-2}$) can be amplified and whether a strong initial magnetic field ($M>0.1$ and $\Lambda>10^{-2}$) can be maintained. The weak-initial-field (WiF) dynamo and strong-initial-field (SiF) dynamo simulations are then classified into four categories based on the resultant behaviour: WiF with a dynamo (e.g., Figure~\ref{fig:time_evolution}a,c), WiF with a dying dynamo (i.e., with a magnetic field that never grows), SiF with a dynamo (e.g., Figure~\ref{fig:time_evolution}a,c), and SiF with a dying dynamo (e.g., Figure~\ref{fig:time_evolution}b).
Most WiF dynamo simulations are run for at least 0.05 magnetic diffusion times $\tau_\eta$ after they reach a saturated state. However, with a WiF dynamo, a metastable state can be found before it reaches the saturated state (Figure~\ref{fig:time_evolution}c). Figure~\ref{fig:time_evolution}a shows a dynamo with the same saturated state growing from a weak-initial-field or a strong-initial-field, without having a long metastable transient. It might take several diffusion times to observe a transition between the metastable and saturated state \cite{p2018}. 
Note that SiF simulations can result in a dying dynamo after a few magnetic diffusion times (e.g., Figure~\ref{fig:time_evolution}b) \cite <e.g.,>{md2009}. 

\begin{figure}
\includegraphics[width=\textwidth]{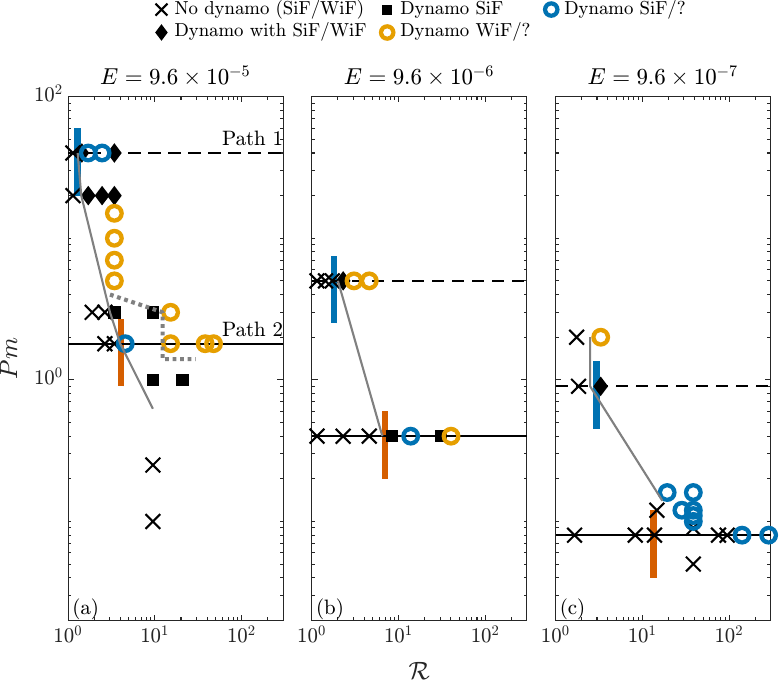}
 \caption{Regime diagram for dynamo action in magnetic Prandtl number $Pm$ and supercriticality $\mathcal{R}$ space for 3 different values of Ekman, $E = 9.8 \times 10^{-5}\mathrm{~(a),~}9.8 \times 10^{-6} \mathrm{~(b)~or~} 9.8 \times 10^{-7}$~(c) (see Figure SI~1 in Supporting information for all other $E$). Crosses denote no dynamo found. Black diamonds denote SiF and WiF dynamo. Black squares denote SiF dynamos and dying WiF dynamos. Empty blue and orange circles denote SiF or WiF dynamo with no constraint on their counterparts (SiF or WiF). Dashed and solid horizontal lines denote $Pm$ values for paths 1 and 2, respectively. Vertical lines (blue and red) show the onset of the dynamo action based on Equation~(\ref{eq:pw1}) (see Figure~\ref{fig:ra_pm_path_rescaled}a). Grey lines highlight the onset of dynamo action. In (a), below the dotted grey line, a dynamo can only grow with a strong-initial-field (above the grey line, a dynamo can grow from a seed field).} 
 \label{fig:rapm}
\end{figure}

In Figure~\ref{fig:rapm}, we show the regime diagrams for dynamo action as a function of the supercriticality $\mathcal{R}$ and $Pm$, for three different Ekman numbers $E=9.8\times 10^{-5}$, $E=9.8\times 10^{-6}$ and $E=9.8\times 10^{-7}$. The equivalent regime diagrams for other Ekman numbers can be found in the Supporting Information (Figure~SI~1). We have classified our simulations into three groups depending on whether the simulations have been initiated with WiF, SiF, or both. Each of the three different groups of simulations can result in an active dynamo or a dying dynamo.

We find that dynamo action sets in at lower levels of supercriticality when the magnetic Prandtl number is higher, in agreement with \citeA{ca2006} and \citeA{blmj2025}. Along path 1 (larger Pm), the onset of dynamo action for both the SiF and WiF dynamos occurs at comparable supercriticalities ($\mathcal{R}_1 = 1.4,\,2.3,\,3.3$ for the three Ekman numbers, respectively). Along path 2, however, the onset depends on both the initial magnetic field state and the Ekman number. In Fig.~\ref{fig:rapm}a,b, two distinct onset thresholds appear: one associated with a strong initial field (SiF), with $\mathcal{R}_{2,SiF}\sim 4$ (a) and $\mathcal{R}_{2,SiF}\sim 10$ (b), and one associated with a weak initial field (WiF), with $\mathcal{R}_{2,WiF}\sim 15$ (a) and $\mathcal{R}_{2,WiF}\sim 40$ (b), for both Ekman number respectively. Thus, the required supercriticality for the SiF dynamo is roughly five times smaller than that for the WiF dynamo, indicating that the flow at SiF onset can maintain a saturated magnetic field but is unable to amplify a seed field (our WiF case).

For $E=9.8 \times 10^{-5}$ (Figure~\ref{fig:rapm}a), we have tested the effect of $Pm$ on the ability to grow a magnetic field from a seed field for a given $\mathcal{R}=3.4$. We have run 8 simulations between $Pm=40$ (path 1) and $Pm=1.8$ (path 2). For $\mathcal{R}=3.4$ and $Pm=1.8$, no dynamo exists regardless of the initial field. When $Pm$ is smaller than 4, the magnetic field cannot exponentially grow from a seed field. 
This small region of parameter space in which a dynamo exists only if initiated with a strong-initial-field has also been observed in simulations with a similar Ekman number and an Earth-like shell geometry \cite{p2018}. This region is related to the two distinct thresholds for the onset of dynamo action at smaller $Pm$.

For $E=9.8 \times 10^{-7}$ (Figure~\ref{fig:rapm}c), dynamo action along path 2 is only found for a strong-initial-field is imposed, within a supercriticality range between $\mathcal{R}_{2,SiF}=94$ and $\mathcal{R}_{2,SiF}=140$. Along this path, the magnetic Prandtl number is 0.08, i.e. much less than 1. In this parameter regime, magnetic field generation depends on small-scale velocity fluctuations, which are ineffective at maintaining large-scale magnetic fields under strong diffusion \cite{t2021}. Thorough exploration of low Ekman and magnetic Prandtl numbers is numerically demanding; at $\mathcal{R}=75,\, 94,\, 140$, the simulations were performed with an imposed azimuthal symmetry $m_p=4$ for some part of the total duration of these simulations \cite{ca2006,sla2025}. As a validation, we carried out a benchmark at $E=9.8\times 10^{-5}$ and $\mathcal{R}_1 = 1.4$, which revealed no substantial differences in the principal diagnostic quantities or in the growth rate of a magnetic field initiated from a seed.

\subsection{Diagnostic parameters along the paths}
\begin{figure}
 \includegraphics[width=\textwidth]{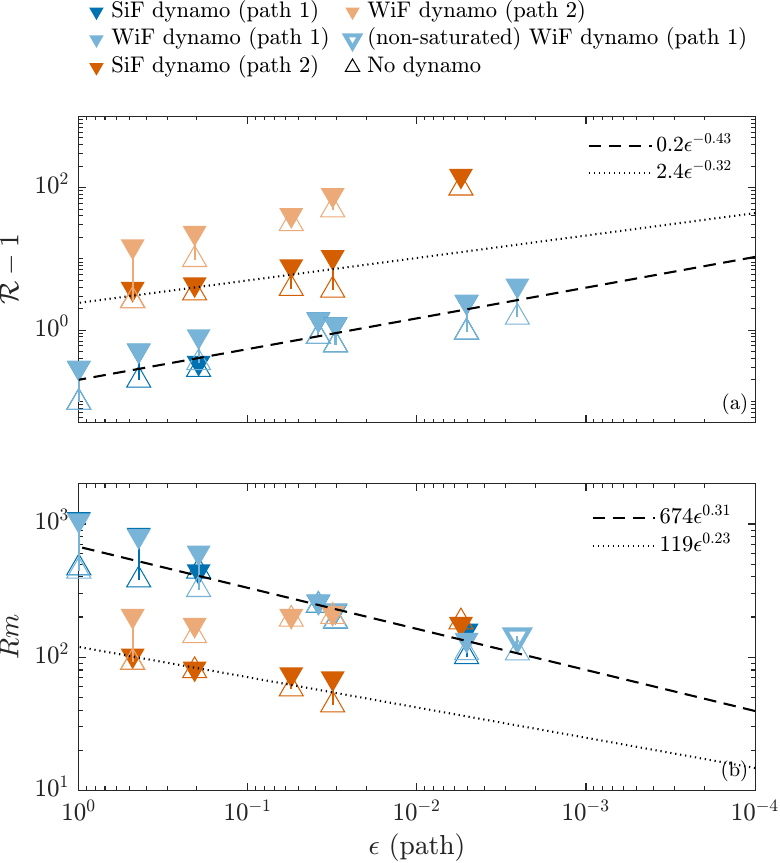}
 \caption{(a) Supercriticality $Ra/Ra_c$ as a function of $\epsilon$ for paths 1 and 2. $\epsilon=1$ is our arbitrary starting point ($Pm=80$ and $E=2 \times 10^{-4}$ for path 1, and $Pm=3$ and $E=2 \times 10^{-4}$ for path 2) and $\epsilon=5\times10^{-10}$ corresponding to the planetesimal parameters ($Pm=10^{-6}$ and $E=10^{-13}$). An upward triangle denotes a dying dynamo, and a downward triangle denotes an active dynamo. Colours represent the initial conditions, a weak-initial-field (WiF) or a strong-initial-field (SiF), and path 1 or 2. Along path 1, dark blue symbols can be hidden by light blue symbols, as the onset of the dynamo behaves in the same way. (b) Magnetic Reynolds number $Rm$ as a function of $\epsilon$ for paths 1 and 2. The downward open triangle denotes the magnetic Reynolds number for the non-saturated dynamo. Note that $E$ and $Pm$ change along the paths.}
\label{fig:ra_pm_path_rescaled}
\end{figure}

Here, we will focus on the supercriticality at the onset of dynamo action and on the main diagnostic quantities at this threshold, for simulations carried out along the two paths introduced in Section~\ref{subsec:path}. In Figure~\ref{fig:ra_pm_path_rescaled}, the values of supercriticality and magnetic Reynolds number at dynamo onset are represented by the interval between the last simulation that does not sustain a dynamo and the first simulation that does, as the Rayleigh number is increased. Equivalently, the true onset of dynamo action lies somewhere between these two limiting cases.

\begin{figure}
 \includegraphics[width=\textwidth]{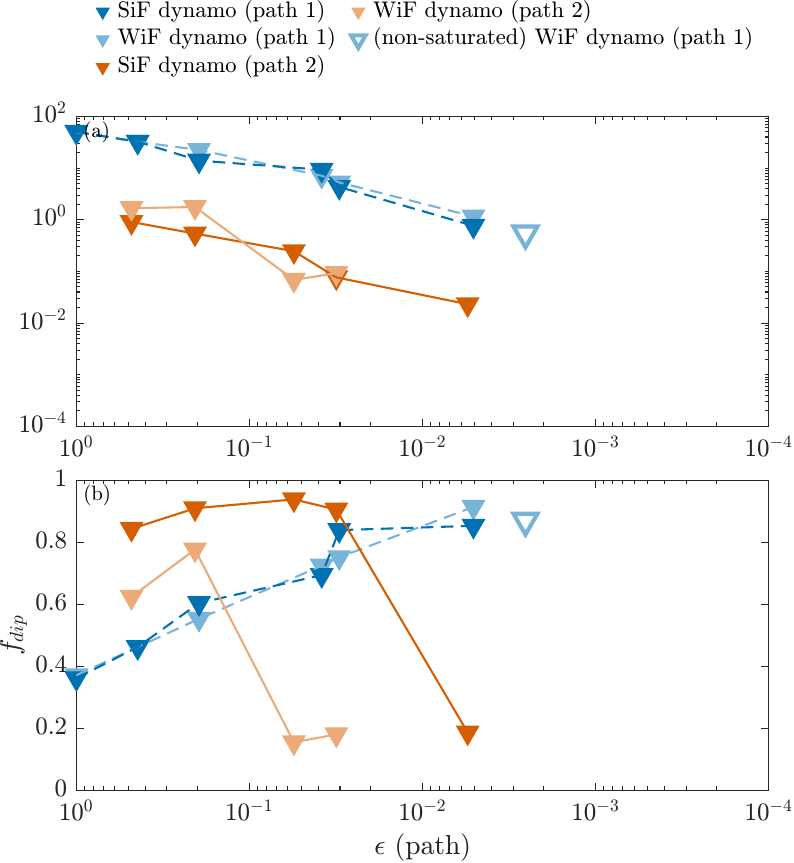}
 \caption{(a) and (b) Elsasser number $\Lambda$ and the dipolar fraction $f_{dip}$ as a function of $\epsilon$ for paths 1 and 2. Colours are similar to those in Figure~\ref{fig:ra_pm_path_rescaled}. Note that $E$ and $Pm$ change along the paths. The downward open triangle denotes a simulation that has not reached a fully saturated state.}
\label{fig:Ell_fdip_path_rescaled}
\end{figure}

Figure~\ref{fig:ra_pm_path_rescaled}a presents the supercriticality as a function of the path parameter $\epsilon$. Along both paths, and regardless of the initial magnetic field, supercriticality increases as the magnetic Prandtl number and Ekman number decrease. Along path 1, the onset of dynamo action increases similarly for both sets of initial conditions (SiF and WiF). In contrast, along path 2, the supercriticality at dynamo onset is highly sensitive to the choice of initial conditions, being significantly higher for a dynamo starting from a weak initial field than for one beginning from an already saturated state. Note that the onset of dynamo action for $\epsilon \sim 0.005$ for path 2 ($E=10^{-6}$ and $Pm=0.08$) is at much higher supercriticality, but corresponds to a multipolar-dominated dynamo rather than the dipolar-dominated dynamos seen along path 2 at higher $\epsilon$ (see Figure~\ref{fig:Ell_fdip_path_rescaled}).

Figure~\ref{fig:ra_pm_path_rescaled}b presents the magnetic Reynolds number along paths 1 and 2 for dynamos initiated with a strong-initial-field (SiF) or a weak-initial-field (WiF). For SiF dynamos, the magnetic Reynolds number decreases as the magnetic Prandtl number and Ekman numbers decrease. 
Since the WiF and SiF dynamos along path 1 have the same onset and saturated states (see Figure~\ref{fig:time_evolution}), the trend in $Rm$ with $\epsilon$ is the same for the WiF and SiF cases. 

For SiF dynamos at smaller $Pm$ (path 2), $Rm$ also monotonically decreases as $\epsilon$ tends to $5 \times 10^{-10}$, except for the lowest value reached ($\epsilon \sim 0.005$). The supercriticality of this simulation at $Pm=0.08$ is also much larger (see Figure~\ref{fig:rapm}c and Figure~\ref{fig:ra_pm_path_rescaled}a). Along path 2, the magnetic Reynolds number of WiF dynamos is about $200$ and is almost independent of the path parameter, while supercriticality increases (Figure~\ref{fig:ra_pm_path_rescaled}a). 

Our simulations span roughly two and a half orders of magnitude in the path parameter, and both supercriticality and magnetic Reynolds number evolve mostly linearly in a log-log space. Therefore, it is appropriate to fit power laws for the evolution of supercriticality and the magnetic Reynolds number along paths 1 ($\propto A_1\epsilon^{p_1}$) and 2 ($\propto A_2\epsilon^{p_2}$) for an initially saturated magnetic field. For $\epsilon = 5 \times 10^{-10}$, all parameters converge for paths 1 and 2; so, both the onset supercriticality and the magnetic Reynolds number should coincide. For the power-law fits, we take the mean between the upper and lower bounds at each $\epsilon$, weighted by the normalised separation of these bounds ($w_{1,2}=\mathrm{lower}/\mathrm{upper}$). We then apply a minimisation routine to determine the minimum of the following multivariable function
\begin{eqnarray}
 \mathcal{F}(A_1,p_1,A_2,p_2)&=&
 w_1(A_1+p_1\,\log_{10}(\epsilon_1)-\log_{10}(y_1))^2 \nonumber \\
 &+&
 w_2(A_2+p_2\,\log_{10}(\epsilon_2)-\log_{10}(y_2))^2 \nonumber \\
 &+&
 \mathcal{P}(10^{A_1}\epsilon_A^{p_1}-10^{A_2}\epsilon_A^{p_2})^2
\end{eqnarray}
where subscripts $1$ and $2$ refer to paths 1 and 2, respectively. The variable $y$ denotes either the supercriticality or the magnetic Reynolds number. The factor $\mathcal{P}$ is a penalty weight introduced to force both power laws to converge to the same value at $\epsilon_A = 5 \times 10^{-10}$. We employ an iterative procedure in which $\mathcal{P}$ is gradually increased until the condition $(10^{A_1}\epsilon_A^{p_1}-10^{A_2}\epsilon_A^{p_2})^2 < 10^{-12}$ is satisfied. This approach has been applied to both the supercriticality and the magnetic Reynolds number.

For the supercriticality, $\mathcal{R}$, in strong-initial-field dynamos with a dipole-dominated configuration (see Figure~\ref{fig:Ell_fdip_path_rescaled} for cases with $f_{dip}\ge0.4$), the optimal power-law fits are
\begin{equation}
\mathcal{R}_1-1 = 0.2\, \epsilon^{-0.43} \quad \mathrm{ and } \quad \mathcal{R}_2-1 = 2.4 \,\epsilon^{-0.32}
 \label{eq:pw1}
\end{equation}
for paths 1 and 2, respectively, with an R-squared value exceeding $93\%$. We fit $\mathcal{R}-1$ rather than $\mathcal{R}$ because $\mathcal{R}$ can be close to unity in our simulations; however, in the regime where $\mathcal{R} \gg 1$, we can safely approximate $\mathcal{R}-1 \approx \mathcal{R}$.

For the magnetic Reynolds number $Rm$ for strong-initial-field dynamos, the best-fit power laws are 
\begin{equation}
Rm_1 = 674 \, \epsilon^{0.31} \quad \mathrm{ and } \quad Rm_2 =119 \, \epsilon^{0.23} 
 \label{eq:pw2}
\end{equation}
for paths 1 and 2, respectively, with an R-squared value above $95\%$. The $Rm$ value for the unsaturated state dynamo at $\epsilon=2 \times10^{-3}$ is not used in the best fit.

Figure~\ref{fig:Ell_fdip_path_rescaled} presents the Elsasser number, $\Lambda$, (Equation~(\ref{eq:ell})) and the dipolar fraction $f_{dip}$ (Equation~(\ref{eq:fdip})) as functions of the path parameter, $\epsilon$, for saturated dynamos. Along path 1, SiF and WiF dynamos converge to the same saturated state. Therefore, both simulations produce the same magnetic field (same $\Lambda$ and $f_{dip}$). For all paths and initial conditions, the Elsasser number decreases with decreasing $\epsilon$ (i.e., with decreasing $E$ and $Pm$). As the Elsasser number becomes smaller than unity along paths 1 and 2, the Lorentz force becomes much smaller than the Coriolis force, leading to a weak-field dynamo \cite{p2018} (see Figure~SI~2 in Supporting Information).
 
Most of the simulations at the onset of dynamo action result in a dipolar or strongly dipolar magnetic field ($f_{\mathrm{dip}}>0.6$). The simulations with the two smallest $\epsilon$ along path 1 are strongly dipolar but have an Elsasser number close to unity, which is a signature for the weak-field dipolar regime as described by \citeA{dormy2025rapidly}, in agreement with the limit found by \citeA{teed2025scaling} (see Figure~\ref{fig:PmE}). For the WiF dynamos on path 2, the onset of dynamo action moves across the transition between the dipolar-dominated to multipolar-dominated regimes as $\epsilon$ decreases. The same behaviour exists for SiF dynamos on path 2 at smaller $\epsilon$ (i.e., smaller $E$ and $Pm$). The supercriticality at the onset of dynamo action for the WiF cases lies between 11 and 24 at $E=4.3\times 10^{-5}$ and between 32 and 42 at $E=9.8\times 10^{-6}$, which are respectively in the dipolar-dominated region and multipolar-dominated region as described in previous simulations at similar Ekman numbers \cite{ca2006,p2018,blmj2025}. For example, the transition occurs at $\mathcal{R}>30$ in the dynamo simulations of \citeA{ca2006} in a spherical shell. We note that the three dynamos yielding a multipolar-dominated magnetic field may lie on a bistable branch as described by \citeA{p2018}, where a multipolar dynamo can be observed before the dipolar/multipolar transition depending on the initial condition. However, we performed only one simulation for each set of parameters, so we did not thoroughly investigate this behaviour.

\subsection{Scaling law for velocity}

\begin{figure}
 \includegraphics[width=\textwidth]{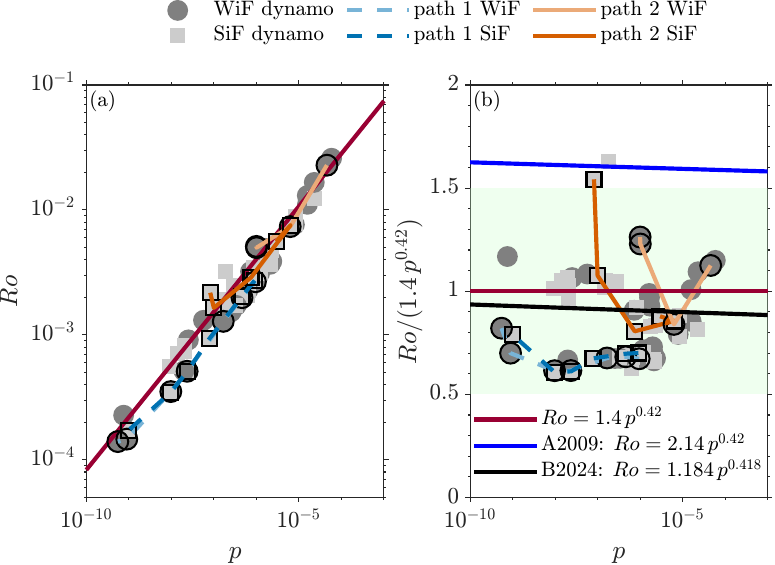}
 \caption{(a) Rossby number $Ro$ as a function of the convective power $p$. Grey squares and circles denote dynamo initiated with a WiF or SiF, respectively. Black edges around squares and circles show the dynamo at the onset of dynamo action along paths 1 and 2 for both WiF and SiF dynamo. Dark red line is the best fit with $Ro =1.4 p^{0.42}$ using dynamos with Nusselt number larger than 3. (b) The ratio between the Rossby number and our best fit as a function of the convective power. Blue line is the scaling law from \citeA{alp2009}. The black line is the scaling law obtained by the best fit of full sphere dynamo simulations \cite{blmj2025} (personal communications). Greenish region denotes the $\pm 50\%$ of the prefactor.}
\label{fig:Ro_p_Ro_p}
\end{figure}

Here, we present the Rossby number (typical velocity normalised by the rotation rate) as a function of the convective power (Figure~\ref{fig:Ro_p_Ro_p}). Our simulations are performed close to the onset of dynamo action, which, at high Ekman and magnetic Prandtl numbers, can lie close to the onset of convection; the Nusselt number (the ratio between the total heat flux and the conductive heat flux) is in several cases only slightly above 1. The scaling law between the Rossby number and the convective power (see Equation~(\ref{eq:rossby})) assumes that the outer core is well-mixed, meaning that the Nusselt number is much larger than unity. When fitting a power law to our data, we exclude simulations with $Nu<3$. We note, however, that including all simulations leads to a best fit that differs only marginally. The best fit obtained from our selected simulations is
\begin{equation}
Ro = (1.4\pm 0.5) p^{0.422 \pm 0.024}
 \label{eq:rossby}
\end{equation}
which is very close to $Ro = 2.14 p^{0.42}$ obtained by \citeA{ca2006} and \citeA{alp2009} for different aspect ratios ($0.05<\chi<0.5$) or $Ro = 1.184 p^{0.418}$ derived from the full-sphere dynamo dataset of \citeA{blmj2025} (personal communications). Note that the prefactor in $Ro = c_1 p^b$ is different if $Ro$ is adimensionzalied using $r$ or $D$ and relates as $c_1=c_1^\star (1-\chi)\left(\frac{1}{(1-\chi^3)}\frac{1}{(1-\chi)^{5}}\right)^b\sim 2.14$ for $c_1^\star = 1.31$ and $\chi=0.35$ for \citeA{alp2009}.

Fig~\ref{fig:Ro_p_Ro_p}b further shows that none of these scaling laws fully captures the variability of the Rossby number in our simulations. Although the exponent is fairly well constrained, thanks to the six-decade span in convective power, the prefactor remains less certain due to the large scatter around its mean value, resulting in an uncertainty of $\pm 50\%$ (our prefactor lies between 0.9 and 2.4). \citeA{ca2006} attempted to improve their fit by adding an extra dependence on $Pm^d$, with $d \sim 0.1$, but this dependence is thought to disappear for $Pm<1$ \cite{ct2004}. Using Equation~(\ref{eq:rossby}) to estimate the magnetic Reynolds number ($Rm = Ro \, Pm/E$) from the predicted convective power of a thermal history model would lead to an uncertainty of about $50\%$, which can be significant when $Rm$ is close to its critical value. 

\subsection{Extrapolation to planetesimals}

\begin{figure}
 \includegraphics[width=\textwidth]{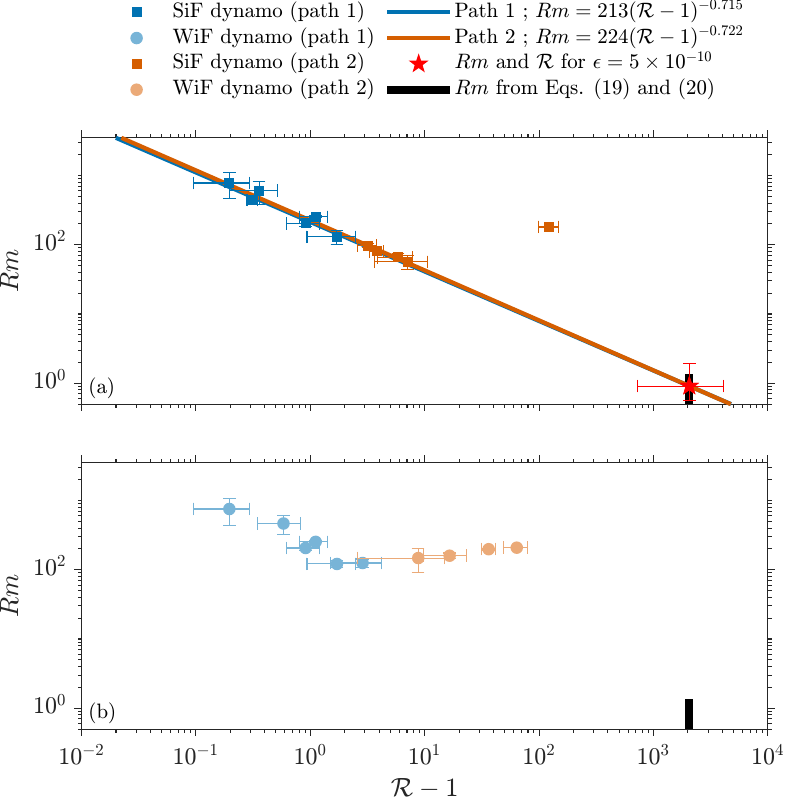}
 \caption{Magnetic Reynolds number $Rm$ as a function of supercriticality $\mathcal{R}-1$ at the onset of dynamo action for SiF dynamos (a) or for WiF dynamos (b). In (a) and (b), squares and circles denote the mean of the lower and upper bounds for the onset of the dynamo for each of paths 1 and 2, using the results from Figure~\ref{fig:ra_pm_path_rescaled}(a and b). Red and blue error bars show the gap between the lower bound (no dynamo) and the upper bound (dynamo). Blue and red lines denote the best fit scaling between $Rm$ and $\mathcal{R}-1$ using the path parameter $\epsilon$. The vertical black lines are the magnetic Reynolds number from Equation~(\ref{eq:rm_MAC}) and (\ref{eq:rm_CIA}), respectively and with $E=10^{-13}$, $Pm=10^{-6}$ and $Pr = 1$ and with $\mathcal{R}=2057$.}
\label{fig:Rm_Ra_extra}
\end{figure}

Although we perform simulations down to a small Ekman number (order $10^{-7}$), our simulations only investigate about 30\% of the path (reaching $\epsilon \sim 2\times 10^{-3}$) towards realistic parameters for planetesimal cores. The supercriticality for the onset of dipolar-dominated dynamos monotonically increases, whereas the magnetic Reynolds number monotonically decreases (see Figure~\ref{fig:ra_pm_path_rescaled}). Therefore, we use the predicted power laws (Equation~(\ref{eq:pw1}) and Equation~(\ref{eq:pw2})) to extrapolate the supercriticality and the magnetic Reynolds number to the end of the path (i.e., $E = 10^{-13}$ and $Pm = 10^{-6}$). Figure~\ref{fig:Rm_Ra_extra} presents the onset of the dynamo action for a strong-initial-field or a weak-initial-field in the $(Rm, \mathcal{R})$ parameter space. At $\epsilon= 5\times 10^{-10}$ (where the paths intersect), both power laws predict a supercriticality of $\sim2000$ and a magnetic Reynolds number of $\sim 1$ for the onset of dynamo action with a pre-existing saturated magnetic field. Note that the scaling laws between the supercriticality and the magnetic Reynolds number are almost identical for both paths (Figure~\ref{fig:Rm_Ra_extra}a). The outlier simulations on path 2 at $E=10^{-6}$ represent the onset of dynamo action in the multipolar region.

One can obtain the magnetic Reynolds number as a function of the supercriticality (or, equivalently, the convective power) \cite{alp2009} using the scaling laws for the velocity (Equation~(\ref{eq:rm_CIA}) or ~(\ref{eq:rm_MAC})). For a 300-km radius core, we will use typical values of $E=10^{-13}$, $Pm=10^{-6}$, and $Pr \sim 1$. At the onset of dynamo action, $\mathcal{R}=2056$, both scaling laws predict $Rm$ close to the value obtained using Equation~(\ref{eq:pw1}) or Equation~(\ref{eq:pw2}). Therefore, these scaling laws seem to be applicable up to the onset of dynamo action, but they do not by themselves determine the critical value of $Rm$.

For a weak-initial-field, the change of regime (dipolar dominated to multipolar dominated) for path 2 prevents the use of a power law to predict the supercriticality and magnetic Reynolds number at the onset of dynamo action at planetesimal conditions (Figure~\ref{fig:Rm_Ra_extra}b).

Our extrapolation of the magnetic Reynolds number at the onset of dynamo action under planetesimal conditions yields a value close to unity, which may seem surprising since commonly cited bounds and critical values are much higher. However, the critical magnetic Reynolds number depends on how the characteristic length and velocity scales are defined. If the length scale is taken as the core radius, the velocity scale can be based on the maximum strain rate $Rm_S$, the maximum velocity $Rm_U$, the root-mean-square (rms) vorticity $Rm_\omega$, or the rms velocity (equivalently, the square root of the kinetic energy) $Rm$. Each of these choices produces a different definition of $Rm$ and thus different lower bounds and optimal thresholds:
\begin{equation}
 Rm =\frac{u L}{\eta}, \quad Rm_S =\frac{e_{\max} L^2}{\eta}, \quad Rm_{U} \frac{U_{\max} L}{\eta}\quad \mathrm{or,}\quad Rm_\omega = \frac{\omega L^2}{\eta}  .
 \end{equation}
Note that it is the magnetic Reynolds number based on the strain rate $Rm_S$ that is equal to  $\pi^2$ \cite{b1958}. Furthermore, when $Rm$ is defined in terms of the rms velocity, no theoretical lower bound exists \cite{p2015,t2021}, and dynamo action can, in principle, occur at arbitrarily small $Rm$, provided the magnetic field varies on spatial scales much larger than those of the flow. A further source of ambiguity arises from the discrepancies between theoretical lower bounds \cite{b1958,c1969,p1977,p1979,chlllj2018}, optimal (or minimal) values for specific flow configurations \cite{w2012,chj2015,chlllj2018,lcll2020}, and numerically determined critical values \cite{ca2006,sla2025}. The smallest theoretical bounds suggest $Rm > \sqrt{3/8}$, whereas a more conservative estimate gives $Rm > 60$ \cite{ca2006,lfncs2022}. At the opposite extreme, low-$Pm$ dynamos are generally expected to require larger magnetic Reynolds numbers \cite{t2021}, yet the combined influence of small-scale turbulence and large-scale rotational constraints can reduce the critical $Rm$, even for dynamos at $Pm \ll 1$. Consequently, our estimated magnetic Reynolds number at the onset of dynamo action may still lie within plausible bounds once we account for the spherical geometry, the low magnetic Prandtl number, the strong rotational constraint, and the presence of an initially strong magnetic field.

\section{Discussion}
\label{sec:discussion}

The onset of dynamo action has previously been extensively investigated using theoretical models and numerical simulations across a wide range of geometries, driving mechanisms, and initial conditions, with applications spanning laboratory experiments and astrophysical and geophysical bodies. Despite this significant amount of work, the critical $Rm$ required for dynamo action in a given configuration remains poorly constrained. Recent studies of thermal evolution models of planetesimal cores \cite{sbn2024,sbnd2025,dodds2021thermal} suggest that the magnetic Reynolds number in such bodies may be smaller than 20 and close to the critical threshold. The existence of a magnetic field for a given thermal evolution is thus highly sensitive to the assumed critical value. Moreover, estimates of the magnetic Reynolds number rely on scaling laws whose prefactors are not well constrained, further increasing this uncertainty.

In this study, we investigated the onset of dynamo action along two paths in $(E, Pm)$ space, both starting at $E = 2 \times 10^{-4}$ with $Pm = 80$ or $Pm = 3$, and ending at $E = 10^{-13}$ and $Pm = 10^{-6}$.  Along these paths, we considered simulations starting from both weak (WiF) and strong (SiF) initial field states and tracked the supercriticality at dynamo onset and three key magnetic diagnostics: the magnetic Reynolds number $Rm$, the Elsasser number $\Lambda$, and the dipolar fraction $f_{\mathrm{dip}}$. For dynamos with a SiF, the supercriticality increases, and the magnetic Reynolds number decreases monotonically when considering only dipole-dominated solutions. Along path 2, for WiF dynamos, the onset of dynamo action first produces a dipole-dominated field, followed by a transition to a multipolar-dominated field as the dipole–multipole boundary is crossed. Similar to \citeA{p2018}, we identified a parameter regime near the minimum $Pm$ for dipole-dominated dynamos in which a dynamo can sustain a pre-existing SiF but cannot exponentially amplify a WiF. This highlights the importance of the initial magnetic field state when assessing dynamo capability close to onset.

Our simulations necessarily investigated only $30\%$ of both paths; therefore, the last simulations along both paths are still far from planetesimal conditions. For both initial conditions, we find that our dynamo simulations are geostrophic at leading order, while viscous forces become increasingly subdominant along the paths. However, for the WiF simulations, both paths evolve through different dynamo states, leading to dipolar or multipolar dynamos. This behaviour prevents extrapolation of the path to planetesimal conditions. For strong-initial-field dynamos, because both paths evolve monotonically in $E$ and $Pm$ and converge toward the same asymptotic values, we were able to derive empirical power-law relationships for the supercriticality and for the magnetic Reynolds number. Extrapolating these trends to planetesimal conditions suggests that dynamo action can occur for supercriticality larger than $10^3$ (corresponding to $Ra \gg 10^{21}$), with a magnetic Reynolds number exceeding unity. We expect that the onset of WiF dynamo action will occur at larger supercriticality and larger $Rm$ than for a strong-initial-field.

The extrapolated critical magnetic Reynolds number obtained here ($Rm \sim 1$) is somewhat smaller than values estimated from previous numerical dynamo simulations \cite{ca2006}, but is comparable to some theoretical estimates \cite{t2021}. The discrepancy may be due to many factors, including the choice of boundary conditions, the method for sampling parameter space, and the influence of an existing saturated magnetic field at the onset. Sustaining an existing magnetic field appears to require lower supercriticality, as it is easier to maintain a magnetic field than to generate one from a WiF. Therefore, a smaller $Rm$ exists at the onset of dynamo action with a SiF than a WiF.

The scaling derived in our study and previous studies \cite{ca2006,alp2009,blmj2025} for the Rossby number and convective power lies between the two classical velocity scaling laws \cite{c2010}. Applying these scaling laws to estimate the magnetic Reynolds number at dynamo onset for planetesimal conditions (at $\mathcal{R} \sim 10^3$ and with $E = 10^{-13}$ and $Pm = 10^{-6}$) yields values comparable to those obtained from our extrapolation. Previous studies have often assumed that $Rm\sim 60$ at dynamo onset, corresponding to a supercriticality $\mathcal{R} > 10^6$, which is much larger than our estimated supercriticality for the onset of dynamo action ($\mathcal{R} \sim 2000$).

\begin{figure}
 \includegraphics[width=\textwidth]{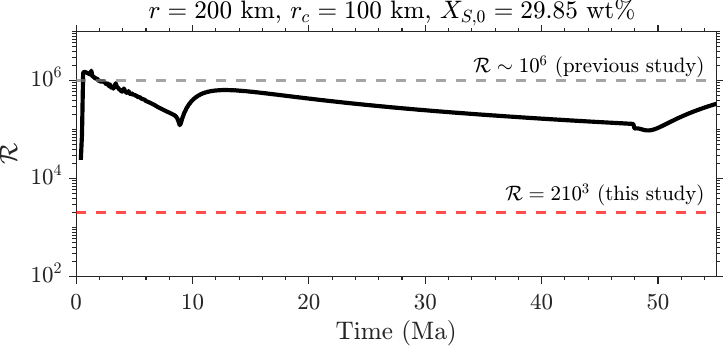}
 \caption{Evolution of the convective supercriticality before the core crystallisation in a 200-km planetesimal radius (with a core radius $r_c = 100$~km). Parameters are identical to the Figure 7 and 8 of \citeA{sbnd2025} (accreted at 0.8 Ma after CAI formation, $^{60}\mathrm{Fe}/^{56}\mathrm{Fe}=10^{-8}$, and $\mathrm{X_{S,0}}=29.85$~wt$\%$), except for the planetesimal and core radius. $Ra_c$ is set to $4.7 \times 10^{17}$. Grey and red dashed lines correspond to the supercriticality criteria based on previous studies and our study respective lily. Details on the thermal evolution model can be found in \citeA{sbnd2025} and  in the Supporting Information (Figure~SI~3).} 
\label{fig:thermalhistory}
\end{figure}

In thermal evolution models of planetesimals, convective supercriticality ($\mathcal{R}$) is obtained directly, whereas the magnetic Reynolds number must be estimated via scaling laws that involve substantial uncertainties. Adopting the planetesimal thermal evolution framework of \citeA{sbnd2025}, the threshold for dynamo onset found here (i.e., $\mathcal{R}\sim 2000$) facilitates the generation of a magnetic field by both extending its active duration and allowing a smaller minimum core radius (Figure~\ref{fig:thermalhistory}). For a large planetesimal with a 500-km radius and a core radius fraction equal to $50\%$ (comparable to that shown in Figure~8 of \citeA{sbnd2025}), the temporal gap between a thermally driven and a compositionally driven dynamo would decrease from $\sim 30$~Myr to about $\sim 7$~Myr. For a smaller planetesimal (with a radius up to 200 km and a core radius fraction of $50\%$), magnetic field generation would be continuous, with no gap between thermal and compositional dynamo generation (Figure~\ref{fig:thermalhistory}). Earlier work has generally found that very small planetesimals (100 km radius or less) cannot sustain a magnetic field unless it is powered by core crystallisation \cite{dodds2021thermal,sbn2024}. In contrast, our results indicate that for small planetesimals (with a core radius of $\sim 40$~km in an 80-km radius planetesimal), thermally driven convection exceeds the dynamo onset criterion for $\mathcal{R}$ by at least an order of magnitude. Therefore, if the magnetic field of the solar nebula or a neighbouring body provides a sufficiently strong initial seed \cite{weiss2021history}, dynamo action can be sustained over longer intervals and in planetesimals with smaller core radii. However, if only a weak seed field is present (with magnetic energy much smaller than the kinetic energy), initiating dynamo action would demand greater convective power, and thus a higher magnetic Reynolds number (relative to previous studies) or an additional energy source (mechanical or compositional).
Our results indicate that, in small bodies, dynamo action can maintain a magnetic field once the convective supercriticality surpasses $10^3$. This finding implies a revision of the minimum core size (and planetesimal size) needed for magnetic field generation and suggests a prolonged lifetime for such fields \cite{bnn2019,dodds2021thermal,sbnd2025,sbn2024}. The thermal and magnetic evolution of planetesimals is highly sensitive to the chosen planetesimal parameters (including the concentration of radioactive isotopes, mantle rheology, core fraction, and the initial sulfur content of the core). A follow-up study will systematically apply this new threshold to reassess the planetesimal properties required to generate magnetic fields and compare this to the meteorite record.

\section*{Data Availability Statement}
A table with input parameters and diagnostic parameters for all 142 simulations is available here \cite{hmdb2026}. Data and Matlab script to reproduce the figures can also be found in \citeA{hmdb2026}. 

\section*{Conflicts of Interest}
The authors declare there are no conflicts of interest for this manuscript.

\section*{Acknowledgements}

We gratefully acknowledge support from the Science and Technology Facilities Council, project references ST/Y001206/1 and ST/Y001796/1. Calculations were performed on the ARCHER2 UK National Supercomputing Service (https://www.archer2.ac.uk) and on the Aire HPC system at the University of Leeds, UK. We also thank Hannah Sanderson for providing the thermal evolution code for planetesimals and for fruitful discussions about the implications of our work on magnetic field generation in planetesimals.



\bibliography{biblio}

@article{wsg2007,
  title={Thermal core--mantle interaction: exploring regimes for ‘locked’dynamo action},
  author={Willis, Ashley P and Sreenivasan, Binod and Gubbins, David},
  journal={Physics of the Earth and Planetary Interiors},
  volume={165},
  number={1-2},
  pages={83--92},
  year={2007},
  publisher={Elsevier}
}

@Article{b1975,
author={Busse, F.H.},
title ={{A model of the geodynamo}},
pages ={437-459},
volume={42},
journal ={Geophysical Journal of the Royal Astronomical Society},
publication-type={J},
year={1975},
}

@article{ct2004,
author={Christensen, Ulrich R. and Tilgner, A.},
journal={Nature},
pages={169-171},
Publication-Type={J},
title={{Power requirement of the geodynamo from ohmic losses in numerical and laboratory dynamos}},
volume={439},
year={2004}}

@article{ca2006,
author={Christensen, Ulrich R. and Aubert, J.},
journal={Geophysical Journal International},
pages={97-114},
Publication-Type={J},
title={{Scaling properties of convection-driven dynamos in rotating spherical shells and application to planetary magnetic fields}},
volume={166},
year={2006}}

@article {rbs2015,
author={Rückriemen, T. and Breuer, D. and Spohn, T.},
title={{The Fe snow regime in Ganymede's core: A deep-seated dynamo below a stable snow zone}},
journal={Journal of Geophysical Research},
issn={2169-9100},
doi={10.1002/2014JE004781},
pages={n/a-n/a},
year={2015},
note={2014JE004781},
}

@article{bwhhk2017,
title={{Paleomagnetic evidence for dynamo activity driven by inward crystallisation of a metallic asteroid}},
author={Bryson, James FJ and Weiss, Benjamin P and Harrison, Richard J and Herrero-Albillos, Julia and Kronast, Florian},
journal={Earth and Planetary Science Letters},
volume={472},
pages={152-163},
year={2017},
publisher={Elsevier}
}

@article{bnn2019,
title = {Constraints on asteroid magnetic field evolution and the radii of meteorite parent bodies from thermal modelling},
author = {Bryson, James FJ and Neufeld, Jerome A and Nimmo, Francis},
journal = {Earth and Planetary Science Letters},
volume = {521},
pages = {68--78},
year = {2019},
publisher = {Elsevier}
}

@article{kv2018,
title = {The thermal evolution of Mercury's Fe--Si core},
author = {Knibbe, Jurri{\"e}n Sebastiaan and van Westrenen, Wim},
journal = {Earth and Planetary Science Letters},
volume = {482},
pages = {147--159},
year = {2018},
publisher = {Elsevier}
}

@article{lfncs2022,
title={Sustaining Earth’s magnetic dynamo},
author={Landeau, Maylis and Fournier, Alexandre and Nataf, Henri-Claude and C{\'e}bron, David and Schaeffer, Nathana{\"e}l},
journal={Nature Reviews Earth \& Environment},
volume={3},
number={4},
pages={255--269},
year={2022},
publisher={Nature Publishing Group UK London}
}

@article{btcsa2023,
title={Onset of convection in rotating spherical shells: Variations with radius ratio},
author={Barik, A and Triana, SA and Calkins, M and Stanley, S and Aurnou, J},
journal={Earth and Space Science},
volume={10},
number={1},
pages={e2022EA002606},
year={2023},
publisher={Wiley Online Library}
}

@article{dsjjc2004,
title={The onset of thermal convection in rotating spherical shells},
author={Dormy, E and Soward, AM and Jones, CA and Jault, D and Cardin, P},
journal={Journal of Fluid Mechanics},
volume={501},
pages={43--70},
year={2004},
publisher={Cambridge University Press}
}

@article{dl2008,
title={Geomagnetism and the dynamo: where do we stand?},
author={Dormy, Emmanuel and Le Mou{\"e}l, Jean-Louis},
journal={Comptes rendus. Physique},
volume={9},
number={7},
pages={711--720},
year={2008}
}

@article{blmj2025,
title={Rapidly rotating early-Earth dynamos in a full-sphere geometry},
author={Burmann, F and Luo, J and Marti, P and Jackson, A},
journal={Geophysical Journal International},
volume={241},
number={1},
pages={715--727},
year={2025},
publisher={Oxford University Press}
}

@article{alp2009,
title={Modelling the palaeo-evolution of the geodynamo},
author={Aubert, Julien and Labrosse, St{\'e}phane and Poitou, Charles},
journal={Geophysical Journal International},
volume={179},
number={3},
pages={1414--1428},
year={2009},
publisher={Blackwell Publishing Ltd Oxford, UK}
}

@article{Mound2017HeatConditions,
title = {{Heat transfer in rapidly rotating convection with heterogeneous thermal boundary conditions}},
year = {2017},
journal = {Journal of Fluid Mechanics},
author = {Mound, Jon E. and Davies, Christopher J.},
pages = {601--629},
volume = {828},
doi = {10.1017/jfm.2017.539},
issn = {14697645}
}

@article{Matsui2016PerformanceModel,
title = {{Performance benchmarks for a next generation numerical dynamo model}},
year = {2016},
journal = {Geochemistry, Geophysics, Geosystems},
author = {Matsui, Hiroaki and Heien, Eric and Aubert, Julien and Aurnou, Jonathan M. and Avery, Margaret and Brown, Ben and Buffett, Bruce A. and Busse, Friedrich and Christensen, Ulrich R. and Davies, Christopher J. and Featherstone, Nicholas and Gastine, Thomas and Glatzmaier, Gary A. and Gubbins, David and Guermond, Jean Luc and Hayashi, Yoshi Yuki and Hollerbach, Rainer and Hwang, Lorraine J. and Jackson, Andrew and Jones, Chris A. and Jiang, Weiyuan and Kellogg, Louise H. and Kuang, Weijia and Landeau, Maylis and Marti, Philippe and Olson, Peter and Ribeiro, Adolfo and Sasaki, Youhei and Schaeffer, Nathanaël and Simitev, Radostin D. and Sheyko, Andrey and Silva, Luis and Stanley, Sabine and Takahashi, Futoshi and Takehiro, Shin Ichi and Wicht, Johannes and Willis, Ashley P.},
number = {5},
month = {5},
pages = {1586--1607},
volume = {17},
publisher = {Blackwell Publishing Ltd},
doi = {10.1002/2015GC006159},
issn = {15252027}
}

@article{jsm2000,
title={The onset of thermal convection in a rapidly rotating sphere},
author={Jones, Chris A and Soward, Andrew M and Mussa, Ali I},
journal={Journal of Fluid Mechanics},
volume={405},
pages={157--179},
year={2000},
publisher={Cambridge University Press}
}

@article{p2018,
title={Systematic parameter study of dynamo bifurcations in geodynamo simulations},
author={Petitdemange, Ludovic},
journal={Physics of the Earth and Planetary Interiors},
volume={277},
pages={113--132},
year={2018},
publisher={Elsevier}
}

@article{md2009,
title={The dynamo bifurcation in rotating spherical shells},
author={Morin, Vincent and Dormy, Emmanuel},
journal={International Journal of Modern Physics B},
volume={23},
number={28n29},
pages={5467--5482},
year={2009},
publisher={World Scientific}
}

@article{b1958,
title={A class of self-sustaining dissipative spherical dynamos},
author={Backus, George},
journal={Annals of Physics},
volume={4},
number={4},
pages={372--447},
year={1958},
publisher={Elsevier}
}

@article{c1969,
title={Th{\'e}orie magnetohydrodynamique de l’effet dynamo},
author={Childress, S},
journal={Report, Department of Mechanics, Faculty of Science, University of Paris},
year={1969}
}

@article{p1979,
title={Necessary conditions for the magnetohydrodynamic dynamo},
author={Proctor, Michael R.E.},
journal={Geophysical \& Astrophysical Fluid Dynamics},
volume={14},
number={1},
pages={127--145},
year={1979},
publisher={Taylor \& Francis}
}

@article{p2015,
title={Energy requirement for a working dynamo},
author={Proctor, Michael R.E.},
journal={Geophysical \& Astrophysical Fluid Dynamics},
volume={109},
number={6},
pages={611--614},
year={2015},
publisher={Taylor \& Francis}
}

@article{chlllj2018,
title={The optimal kinematic dynamo driven by steady flows in a sphere},
author={Chen, Long and Herreman, Wietze and Li, Kuan and Livermore, Philip W and Luo, JW and Jackson, Andrew},
journal={Journal of Fluid Mechanics},
volume={839},
pages={1--32},
year={2018},
publisher={Cambridge University Press}
}

@article{p1977,
title={On Backus' necessary condition for dynamo action in a conducting sphere},
author={Proctor, Michael R.E.},
journal={Geophysical \& Astrophysical Fluid Dynamics},
volume={9},
number={1},
pages={89--93},
year={1977},
publisher={Taylor \& Francis}
}

@article{lcll2020,
title={Optimal kinematic dynamos in a sphere},
author={Luo, Jiawen and Chen, Long and Li, Kuan and Jackson, Andrew},
journal={Proceedings of the Royal Society A},
volume={476},
number={2233},
pages={20190675},
year={2020},
publisher={The Royal Society Publishing}
}

@article{chj2015,
title={Optimal dynamo action by steady flows confined to a cube},
author={Chen, Long and Herreman, Wietze and Jackson, Andrew},
journal={Journal of Fluid Mechanics},
volume={783},
pages={23--45},
year={2015},
publisher={Cambridge University Press}
}

@article{sbnd2025,
title = {Unlocking planetesimal magnetic field histories: A refined, versatile model for thermal evolution and dynamo generation},
journal = {Icarus},
volume = {425},
pages = {116323},
year = {2025},
issn = {0019-1035},
doi = {10.1016/j.icarus.2024.116323},
author = {Hannah R. Sanderson and James F.J. Bryson and Claire I.O. Nichols and Christopher J. Davies}
}

@article{weiss2008magnetism,
  title={Magnetism on the angrite parent body and the early differentiation of planetesimals},
  author={Weiss, Benjamin P and Berdahl, James S and Elkins-Tanton, Linda and Stanley, Sabine and Lima, Eduardo A and Carporzen, Laurent},
  journal={Science},
  volume={322},
  number={5902},
  pages={713--716},
  year={2008},
  publisher={American Association for the Advancement of Science}
}

@article{carporzen2011magnetic,
  title={Magnetic evidence for a partially differentiated carbonaceous chondrite parent body},
  author={Carporzen, Laurent and Weiss, Benjamin P and Elkins-Tanton, Linda T and Shuster, David L and Ebel, Denton and Gattacceca, J{\'e}r{\^o}me},
  journal={Proceedings of the National Academy of Sciences},
  volume={108},
  number={16},
  pages={6386--6389},
  year={2011},
  publisher={National Academy of Sciences}
}

@article{tarduno2012evidence,
  title={Evidence for a dynamo in the main group pallasite parent body},
  author={Tarduno, John A and Cottrell, Rory D and Nimmo, Francis and Hopkins, Julianna and Voronov, Julia and Erickson, Austen and Blackman, Eric and Scott, Edward RD and McKinley, Robert},
  journal={Science},
  volume={338},
  number={6109},
  pages={939--942},
  year={2012},
  publisher={American Association for the Advancement of Science}
}

@article{weiss2021history,
  title={History of the solar nebula from meteorite paleomagnetism},
  author={Weiss, Benjamin P and Bai, Xue-Ning and Fu, Roger R},
  journal={Science Advances},
  volume={7},
  number={1},
  pages={eaba5967},
  year={2021},
  publisher={American Association for the Advancement of Science}
}

@article{t2021,
  title={The turbulent dynamo},
  author={Tobias, SM},
  journal={Journal of fluid mechanics},
  volume={912},
  pages={P1},
  year={2021},
  publisher={Cambridge University Press}
}

@article{w2012,
  title={Optimization of the magnetic dynamo},
  author={Willis, Ashley P},
  journal={Physical review letters},
  volume={109},
  number={25},
  pages={251101},
  year={2012},
  publisher={APS}
}

@article{d2013,
  title={Scaling laws for planetary dynamos},
  author={Davidson, PA},
  journal={Geophysical Journal International},
  volume={195},
  number={1},
  pages={67--74},
  year={2013},
  publisher={Oxford University Press}
}

@article{c2010,
  title={Dynamo scaling laws and applications to the planets},
  author={Christensen, Ulrich R},
  journal={Space science reviews},
  volume={152},
  number={1},
  pages={565--590},
  year={2010},
  publisher={Springer}
}

@article{sla2025,
  title={Energetically expensive dynamo action in Earth’s basal magma ocean},
  author={Schaeffer, Nathana{\"e}l and Labrosse, St{\'e}phane and Aurnou, Jonathan M},
  journal={Proceedings of the National Academy of Sciences},
  volume={122},
  number={45},
  pages={e2507575122},
  year={2025},
  publisher={National Academy of Sciences}
}

@article{agf2017,
  title={Spherical convective dynamos in the rapidly rotating asymptotic regime},
  author={Aubert, Julien and Gastine, Thomas and Fournier, Alexandre},
  journal={Journal of Fluid Mechanics},
  volume={813},
  pages={558--593},
  year={2017},
  publisher={Cambridge University Press}
}

@article{dodds2021thermal,
  title={The thermal evolution of planetesimals during accretion and differentiation: Consequences for dynamo generation by thermally-driven convection},
  author={Dodds, KH and Bryson, JFJ and Neufeld, JA and Harrison, Richard J},
  journal={Journal of Geophysical Research: Planets},
  volume={126},
  number={3},
  pages={e2020JE006704},
  year={2021},
  publisher={Wiley Online Library}
}

@article{clns2019,
  title={Precessing spherical shells: flows, dissipation, dynamo and the lunar core},
  author={C{\'e}bron, David and Laguerre, Rapha{\"e}l and Noir, Jerome and Schaeffer, Nathana{\"e}l},
  journal={Geophysical Journal International},
  volume={219},
  number={Supplement\_1},
  pages={S34--S57},
  year={2019},
  publisher={Oxford University Press}
}

@article{la2011,
  title={Equatorially asymmetric convection inducing a hemispherical magnetic field in rotating spheres and implications for the past martian dynamo},
  author={Landeau, Maylis and Aubert, Julien},
  journal={Physics of the Earth and planetary interiors},
  volume={185},
  number={3-4},
  pages={61--73},
  year={2011},
  publisher={Elsevier}
}

@article{dormy2025rapidly,
  title={Rapidly rotating magnetohydrodynamics and the geodynamo},
  author={Dormy, Emmanuel},
  journal={Annual Review of Fluid Mechanics},
  volume={57},
  pages={335--362},
  year={2025},
  publisher={Annual Reviews}
}

@article{d2016,
  title={Strong-field spherical dynamos},
  author={Dormy, Emmanuel},
  journal={Journal of Fluid Mechanics},
  volume={789},
  pages={500--513},
  year={2016},
  publisher={Cambridge University Press}
}

@article{a2019,
  title={Approaching Earth’s core conditions in high-resolution geodynamo simulations},
  author={Aubert, Julien},
  journal={Geophysical Journal International},
  volume={219},
  number={Supplement\_1},
  pages={S137--S151},
  year={2019},
  publisher={Oxford University Press}
}

@article{teed2025scaling,
  title={Scaling of strong-field spherical dynamos},
  author={Teed, Robert J and Dormy, Emmanuel},
  journal={Geophysical Research Letters},
  volume={52},
  number={20},
  pages={e2025GL118078},
  year={2025},
  publisher={Wiley Online Library}
}

@article{zr2022,
  title={Thermal evolution and magnetic history of rocky planets},
  author={Zhang, Jisheng and Rogers, Leslie A},
  journal={The Astrophysical Journal},
  volume={938},
  number={2},
  pages={131},
  year={2022},
  publisher={The American Astronomical Society}
}

@article{s2010b,
  title={Planetary magnetic fields: Achievements and prospects},
  author={Stevenson, David J},
  journal={Space science reviews},
  volume={152},
  number={1},
  pages={651--664},
  year={2010},
  publisher={Springer}
}

@article{scheinberg2015magnetic,
  title={Magnetic field generation in the lunar core: The role of inner core growth},
  author={Scheinberg, A and Soderlund, KM and Schubert, G},
  journal={Icarus},
  volume={254},
  pages={62--71},
  year={2015},
  publisher={Elsevier}
}

@article{weiss2010paleomagnetic,
  title={Paleomagnetic records of meteorites and early planetesimal differentiation},
  author={Weiss, Benjamin P and Gattacceca, J{\'e}r{\^o}me and Stanley, Sabine and Rochette, Pierre and Christensen, Ulrich R},
  journal={Space Science Reviews},
  volume={152},
  number={1},
  pages={341--390},
  year={2010},
  publisher={Springer}
}

@book{chandrasekhar2013hydrodynamic,
  title={Hydrodynamic and hydromagnetic stability},
  author={Chandrasekhar, Subrahmanyan},
  year={2013},
  publisher={Courier Corporation}
}

@article{hmdb2026,
title = {Dataset for ``Onset of dynamo action in planetesimals''},
author = {Huguet, Ludovic and Mound, Jonathan E. and Davies, Christopher J. and Bryson, F.J. James},
year = {2026},
journal = {},
doi = {10.6084/m9.figshare.32171925.v1}
}

@article{sbn2024,
  title={Early and elongated epochs of planetesimal dynamo generation},
  author={Sanderson, Hannah R and Bryson, James FJ and Nichols, Claire IO},
  journal={Earth and Planetary Science Letters},
  volume={648},
  pages={119083},
  year={2024},
  publisher={Elsevier}
}

\end{document}